\documentclass[conference]{IEEEtran}
  \usepackage{algorithm}
  \usepackage[noend]{algpseudocode}
  \usepackage{varwidth}
\usepackage{tabularx}
  \ifCLASSINFOpdf
    \usepackage[pdftex]{graphicx}
  \else
  \fi

  \usepackage{amssymb}

\usepackage[switch]{lineno}
\usepackage{xcolor,colortbl}
\usepackage{subcaption}

  \usepackage{color}
  
\usepackage{enumitem}
\usepackage{amsmath}

\definecolor{lightgreen}{rgb}{0.56, 0.93, 0.56}
\definecolor{indianred}{rgb}{1, 0.4, 0.4}

\newcommand{\up}{\cellcolor{lightgreen}}
\newcommand{\down}{\cellcolor{indianred}}

\begin{document}
  \IEEEoverridecommandlockouts
  
  
  %
  \title{On Design of Problem Token Questions in Quality of Experience Surveys}

  \author{
	\IEEEauthorblockN{
    Jayant Gupchup\IEEEauthorrefmark{1},
    Ebrahim Beyrami\IEEEauthorrefmark{1},
    Martin Ellis\IEEEauthorrefmark{1},
		Yasaman Hosseinkashi\IEEEauthorrefmark{1},
		Sam Johnson\IEEEauthorrefmark{1}\IEEEauthorrefmark{2},
		Ross Cutler\IEEEauthorrefmark{1}
	}
	\IEEEauthorblockA{
		\IEEEauthorrefmark{1}Microsoft Corporation, \{jayagup, ebbeyram, maellis, yahossei, sajohnso, rcutler\} @microsoft.com
	}
	\IEEEauthorblockA{
		\IEEEauthorrefmark{2}currently affiliated with Facebook, sam.johnson@fb.com
	}
}

  \maketitle

  \begin{abstract}
  User surveys for Quality of Experience (QoE) are a critical source of information. In addition to the
  common ``star rating'' used to estimate Mean Opinion Score (MOS), more detailed survey questions (problem
  tokens) about specific areas provide valuable insight into the factors impacting QoE. This paper
  explores two aspects of the problem token questionnaire design. First, we study the bias introduced
  by fixed question order, and second, we study the challenge of selecting a subset of questions to keep
  the token set small. Based on 900,000 calls gathered using a randomized controlled experiment from a
  live system, we find that the order bias can be significantly reduced by randomizing the display
  order of tokens. The difference in response rate varies based on token position and display design.
  It is worth noting that the users respond to the randomized-order variant at levels that are comparable
  to the fixed-order variant. The effective selection  of a subset of token questions is achieved by
  extracting tokens that provide the highest information gain over user ratings. This selection is known
  to be in the class of NP-hard problems. We apply a well-known greedy submodular maximization method
  on our dataset to capture 94\% of the information using just 30\% of the questions.
  \end{abstract}

  
  %
  \IEEEpeerreviewmaketitle

\newcommand{\sectionspace}{\vspace{-1mm}}

\section{Introduction}
\label{s:introduction}
\sectionspace
Several Internet telephony applications employ end-of-call user surveys to gather data on in-call QoE ~\cite{via-jiang-2016}. 
In addition to the five star rating (MOS~\cite{itut-p.800.1}), the percentage of calls rated 1 or 2 (poor call rate, or PCR)
is often tracked as a measure of media quality.
Previous studies have shown the value of combining PCR with an additional problem token questionnaire (PTQ) to gather detailed insights ~\cite{blind}.
The UI design of the PTQ used in this study is provided in ~\cite{blind}.

The range of questions used to capture these detailed problem areas has been studied in depth ~\cite{itut-p.800, itut-p.910}. 
However, to the best of our knowledge, the impact of presentation order on response rate has not been studied in a live, deployed system.
Our work is motivated by the practical challenges faced in analyzing questionnaire data.
For example, our analysis showed that the contribution of one-way audio (one side can hear, but the other side cannot) to PCR dwarfed the other areas by a factor of two on mobile platforms.
Further investigation, showed that the most important factor for this gap was the display order of questions.
The `no sound' token was placed at the top, and users were $40\%$ more likely to select this area purely due to its position in the survey.
While one-way audio remains one of the top problem areas, we found that after randomization of the order, other impediments such as audio distortion 
and poor image quality occurred at comparable levels.

In mobile environments, the screen size is limited, so if designers want to avoid using a scrollbar, the number of questions needs to be kept small.
The key question is how to minimize the number of questions, while maximizing their power in explaining PCR. Moreover, studies have shown the benefit
of shortening surveys (without losing information) for improving response rate and improved data quality \cite{allen2016impact}.
%
Identifying this subset of questions belongs to a class of NP-hard problems ~\cite{guyon2003introduction}.
In order to solve this, we follow the lead of Krause et al., leveraging the fact that information gain is a submodular function, and can be optimized using provable greedy approaches ~\cite{krause2014submodular, nushi2016learning}.
As shown in Section \ref{s:token-selection}, this approach maps well to the our problem.
 The main contributions of this paper are:
\noindent
\begin{enumerate}[leftmargin=*]
\item{Results of a large scale randomized, controlled experiment in a live VoIP system to show the bias introduced by fixed order questions.}
\item{Provide an efficient solution to select a subset of tokens that maximizes information and minimizes correlation.}
\end{enumerate}

\section{Related Work}
\label{s:related-work}
\sectionspace
There is a rich area of research and practice in general survey design, validation and question order ~\cite{groves2011survey, mcfarland1981effects}. 
Factor analysis is commonly employed to analyze surveys with the number of factors being smaller than the number of questions. ~\cite{weintraub2009validation}.
There are many standards for subjective audio and video quality surveys, such as the ITU standards ~\cite{itut-p.831, itut-p.800, itut-p.910}.
In this paper, our goal is not to replace these surveys, but instead to improve their utility by providing recommendations
for presentation order, and a methodology to select a subset of informative questions.
The selection of a subset of correlated random variables for maximizing the information gain has been studied in detail ~\cite{krause2014submodular, nushi2016learning} -
this paper focuses on the application of these methods for QoE problem area surveys.

\section{Impact of Question Display Order}
\label{s:displayorder}
\sectionspace
We study the impact of question order on our PTQ using a randomized controlled experiment.
The control population was shown the original questionnaire with fixed token order. For video calls, the audio tokens were 
always shown on the left while the video tokens were always shown on the right. The treatment population was shown the 
questions in randomized order. For video calls in the treatment population, the position of the audio and video panels 
(left/right) were selected at random. The details of the experiment are below:
\noindent
\vspace{-1mm}
\begin{itemize}[leftmargin=*]
\item{One-to-one calls that included both audio-only and video VoIP calls on desktop platforms.}
\item{The control and treatment group each contained 450,000 calls spanning over 100,000 unique users.}
\end{itemize}

\noindent
The experiment aimed to answer the following questions:
\noindent
\vspace{-1mm}
\begin{enumerate}[leftmargin=*]
\item{Is there a change in the percentage of questionnaire responses?}
\item{What is the change in response rate of the individual audio and video tokens?}
\end{enumerate}

We present all results using relative measures due to commercial confidentiality.

\subsection{Overall Questionnaire Response Rate}
\label{ss:tok-rr}
\sectionspace
We wanted to understand if randomizing the questions impacts the overall response rate of the PTQ.
A user is said to respond to a PTQ if any token selection is made. We considered audio-only and video population.
The differences in overall token response rate between the control group and the treatment group for these segments is shown in Table \ref{tab:tok-rr}.
From a statistical perspective, there is no change in the percentage of responders in the audio population, but there is a change in the video 
population at the $99\%$ significance level (i.e. $p-value < 0.01$).
However, we find that the relative difference of $1.6\%$ for video surveys is small enough that the benefits 
(Section \ref{ss:tok-rr}) of the randomized questionnaire outweigh the minor reduction in response rate.

\setlength{\textfloatsep}{10pt}
\begin{table}[t]
\caption{The difference in overall response rate in tokens between control group and treatment group}
\label{tab:tok-rr}
\centering
\begin{tabular}{lcc}
\textbf{Population} & \textbf{Relative Delta} & \textbf{p-value} \\
\hline
Audio-only & $-1.38\%$ & $0.072$ \\
Video & $-1.62\%$ & $0.001$ \\
\hline
\end{tabular}
\end{table}

\subsection{Response Rate of Individual Questions}
\label{ss:tok-rr}
\sectionspace
The difference in response rates of individual tokens between the fixed order and randomized order for video, desktop tokens is shown in Table \ref{tab:tok-av-rr}. Negative sign (red) 
indicates that the response rate went down whereas a positive sign (green) indicates the response rate went up. Although a change in the response rate 
was expected, there are some significant insights from these results: 
\begin{enumerate}[leftmargin=*]
\item{The response rate of the top token is dramatically impacted for audio and video tokens. The decrease in response rate between 
the two variants is greater than $20\%$. This shows the propensity for selecting the top token.}
\item{For audio, the response rates of the top four tokens decreased dramatically, However, for video the response rates of the 
bottom four tokens increased. This shows that panel position (left, right) impacts response rate.}
\end{enumerate}

The impact of panel position is even more pronounced in mobile environments. In mobile, we found the average response rate for 
tokens that require users to scroll was $49\%$ lower. These results motivate the need for showing a small set of informative tokens to ensure 
that we do not lose the user’s attention while responding to the questionnaire.

\setlength{\textfloatsep}{10pt}
\begin{table}[t]
\caption{The difference in response rate of individual tokens for fixed vs.~randomized display order in video desktop calls.}
\label{tab:tok-av-rr}
\begin{tabularx}{0.8\linewidth}{lcc}
\hline
\textbf{Audio problem Token} & \textbf{Relative delta} & \textbf{p-value} \\
I could not hear any sound & \down $-26.7\%$ & \down $\leq 2e^{-29}$ \\
The other side could not hear any sound & \down $-12.5\%$ & \down $2e^{-23}$ \\
I heard echo in the call & \down $-12.7\%$ & \down $6e^{-24}$ \\
I heard noise in the call & \down $-9.5\%$ & \down $3e^{-18}$ \\
Volume was low & $-3.8\%$ & $0.01$ \\
The call ended unexpectedly &  $-2.4\%$ & $0.10$ \\
{\scriptsize Speech was not natural or sounded distorted} & \up $+3.6\%$ & \up $0.00$ \\
We kept interrupting each other & \down $-1.7\%$ & \down $0.22$ \\
\hline
\textbf{Video problem Token} & \textbf{Relative delta} & \textbf{p-value} \\
I could not see any video & \down $-20.4\%$ & \down $\leq 2e^{-29}$ \\
The other side could not see my video &  $-1.9\%$ &  $0.39$ \\
Image quality was poor & $-2.2\%$ &  $0.06$ \\
Video kept freezing & \up $+10.1\%$ & \up $3e^{-16}$ \\
Video stopped unexpectedly & \up $+28.0\%$ & \up $4e^{-45}$ \\
The other side was too dark &  \up $+25.2\%$ & \up $8e^{-21}$ \\
Video was ahead or behind audio & \up $+25.2\%$ & \up $5e^{-39}$ \\
\hline

\end{tabularx}
\end{table}

\section{Token Subset Selection}
\label{s:token-selection}
\sectionspace
The question we are trying to address is the following: Given a limited budget of questions, $k$, is there a systematic process of selecting the questions 
to maximize information? In this paper, we propose the selection of tokens by applying the algorithm described by Nushi et al. \cite{nushi2016learning}. 
An overview of this approach is provided next.

\subsection{Information Gain and Submodular Function Optimization}
\label{ss:submodular}
\sectionspace
Information gain ($IG$) captures the amount of information “shared” between two random variables. Mathematically, $IG$ between 
variables $PC$ and $T$, is defined as $IG [PC;T]=H[PC]-H[PC \mid T]$ ; where $H$ represents the entropy (uncertainty) of a random variable. 
It should be clear that $IG$ has the property of monotonicity ~\cite{cover2012elements}. We can easily see that for any two sets of random variables,
$T_1,T_2 \colon T_1 \subset T_2, IG[PC;T_1] \geq IG[PC;T_2]$. Building on the monotonicity property and borrowing notation from [4], IG also has a 
“diminishing returns” property. The incremental information gain obtained by adding a new element to a subset is higher than the 
incremental information gain obtained by adding a new element to its superset. Mathematically, if we consider two set of token variables $T_1,T_2$ 
from the universe of problem tokens, $T$, such that $T_1 \subset T_2 \subset T$, and we consider a token $e:e \notin T_1$ and $e \notin T_2$ then 
the principle of diminishing returns property is shown in the equation below:

\begin{equation}
\begin{split}
IG[PC; \{T_1  \cup \{e\}\}] - IG[PC;T_1] \geq \\
 IG[PC;\{T_2  \cup \{e\}\}] - IG[PC;T_2]
\end{split}
\end{equation}

In other words, the marginal benefit of reducing the uncertainty in $PCR$ by adding a new token to a smaller set is higher than any superset. 
This property is known as “submodularity”. ~\cite{krause2014submodular} Krause et al. show that interaction effects (e.g., mutual exclusion) between 
variables can result in $IG$ to not be strictly submodular. However, we do not see such interaction effects for our token dataset, and therefore no 
violation of submodularity of information gain. 

The optimization of submodular functions is a known NP-hard problem, Nemhauser et al. [3, 4] provide a greedy algorithm to solve this problem that is $63\%$ 
of the computationally expensive and exhaustive solution. The algorithm is iterative and fairly straightforward: In iteration 0, we start with an empty set, $T_0$. 
At every iteration $i$, we add the token, $t$, that maximizes the discrete derivative of information gain and set $T_i$ is constructed using the equation:
\begin{equation}
T_i=T_{i-1} \cup { \arg\max_{t} IG[PC;(T_{i-1} \cup {t})] }
\end{equation}

where $t \in T \setminus T_{i-1}$ and $IG[PC;(T_{i-1} \cup {t})]$ represents the information gain of PC by jointly considering the tokens $(T_{i-1} \cup {t})$ for 
all candidate tokens t. Note that iteration 1 picks the token that provides the highest univariate information gain. In subsequent iterations, this method 
selects tokens that provide the information not already captured by the existing token set. By design, this results in selection of the least correlated tokens at every iteration. 
We view this method of selecting tokens as maximizing the return of information for tokens shown (hereafter referred to as \emph{RITS}).

\setlength{\belowcaptionskip}{0pt}
\begin{figure}[t]
  \centering
  \begin{subfigure}[b]{0.45\linewidth}
    \centering
    \includegraphics[width=\textwidth]{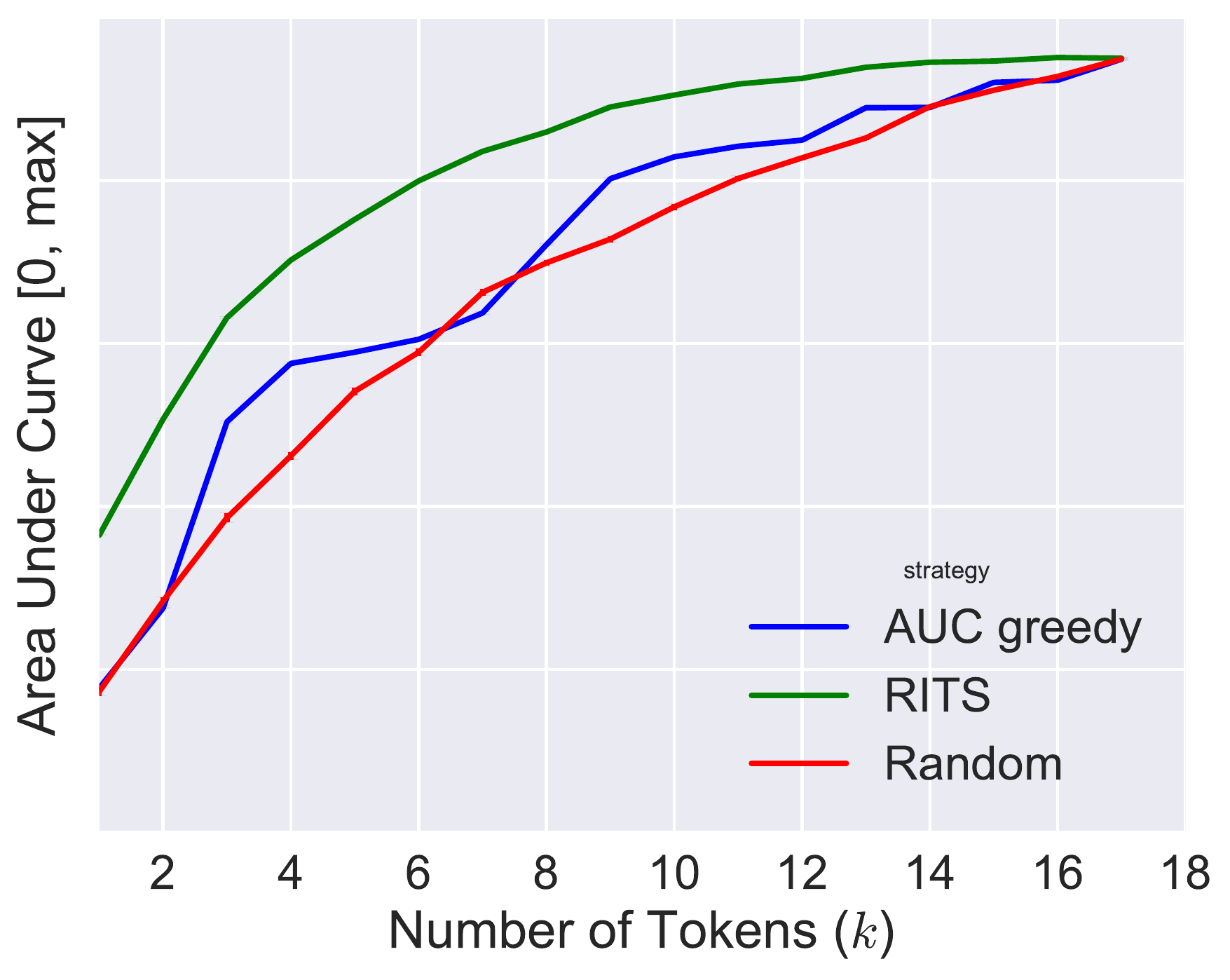}
  \end{subfigure}
  \begin{subfigure}[b]{0.45\linewidth}
    \includegraphics[width=\textwidth]{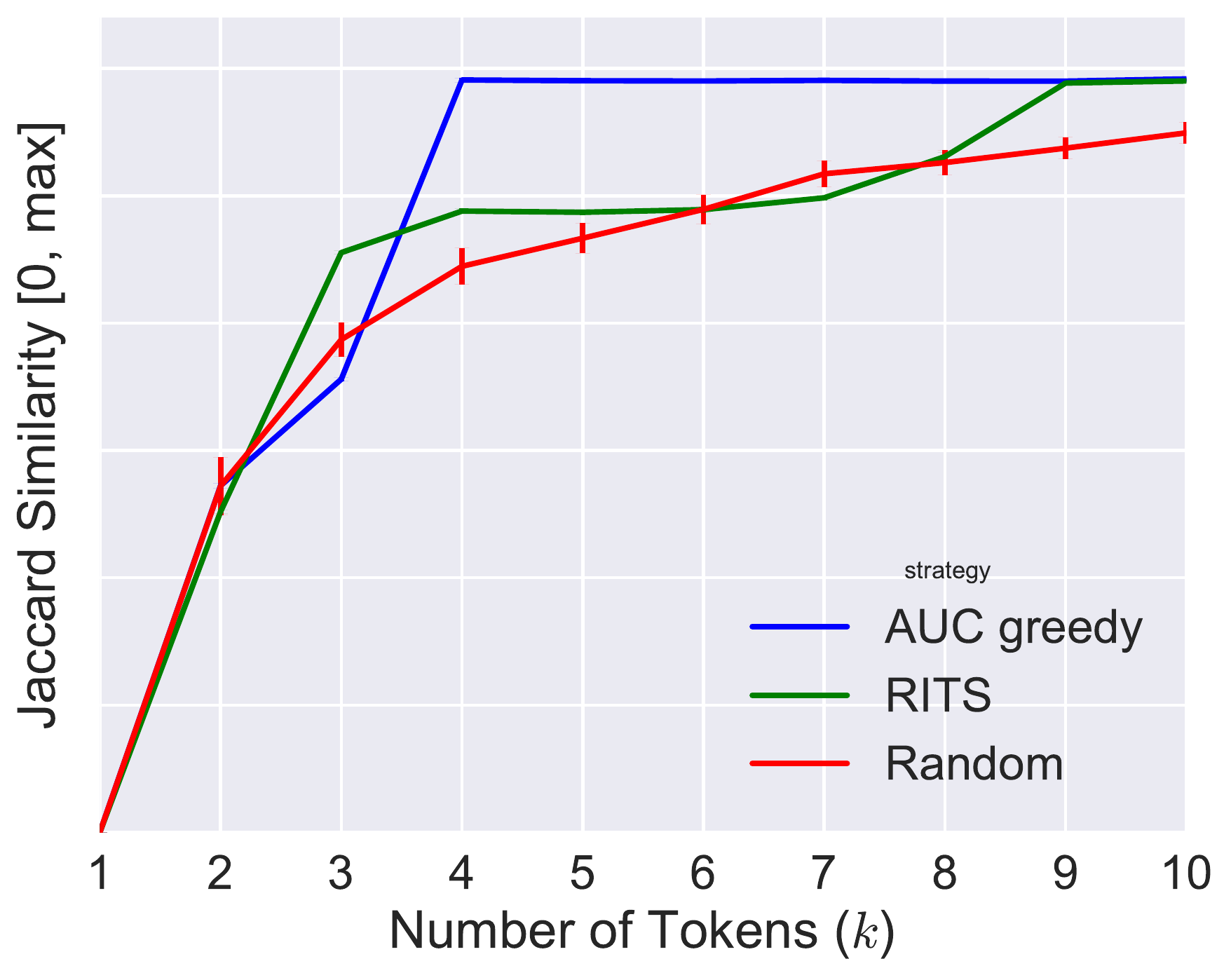}
  \end{subfigure}
  \caption{
  Relative AUC performance of different strategies in selecting tokens is shown on left while  
 Jaccard similarity scores are shown on right. Note: Scales removed for confidentiality. }
  \label{fig:perf}
\end{figure}

\subsection{Evaluation of RITS}
\label{ss:rits}
\sectionspace
Data gathered from the treatment (randomized PTQ) population of the experiment was used for evaluation.
The RITS method was evaluated using the following quantitative metrics:
\noindent
\vspace{-2mm}
\begin{enumerate}[leftmargin=*]
\item{AUC: Area under an ROC curve; \cite{datamining-tan-2006}}
\item{JS: Jaccard Similarity Coefficient \cite{datamining-tan-2006}}
\end{enumerate}

While AUC measures the ability of the token set to discriminate between a good call and a poor call, Jaccard similarity measures the pair-wise degree of overlap between the tokens.
The ideal token set has an AUC close to 1 and a JS score of 0.
Uncorrelatedness (i.e. a low JS score) is important when breaking down an overall quality metric into distinct factors. For a given token count, $k$, we studied and compared the performance of RITS with the following approaches:
\noindent
\vspace{-1mm}
\begin{enumerate}[leftmargin=*]
\item{Random: Select a random subset of tokens.}
\item{AUC-Greedy: Select tokens sorted in descending order of their univariate AUC.}
\end{enumerate}
In our evaluation, we used the random forest implementation from the Python scikit-learn library \cite{pedregosa2011scikit} to obtain the classification boundary 
with default settings. The error bars were obtained using 100 independent runs of train/test splits.

\subsection{Results using RITS}
\label{ss:rits_results}
\sectionspace
The AUC and JS scores of the different token subset selection strategies are shown in Figure \ref{fig:perf}. Note that we represent the scale in terms of the
maximum values obtained for our dataset, and hide the labels for corporate confidentiality. This allows for relative comparison between the selection strategies. 
The RITS method significantly outperforms AUC-greedy and random method for all $k$ in-terms of the AUC criterion. The shape of the RITS curve highlights the "Diminishing" returns property. 
The RITS method also has significantly lower JS score compared to the AUC greedy method for $k>4$. This is because RITS is designed to find the 
tokens that provide information that is not already covered by the existing set of tokens. Since the AUC-greedy method does not consider correlation, it performs poorly on the 
Jaccard similarity measure. Using the RITS method, the first five tokens capture $94\%$ of the total information content in our dataset.


\section{Summary}
\label{s:summary}
\sectionspace
In this paper, we studied two aspects of the problem token questionnaire design for VoIP applications - display order and token subset selection. 
Based on over 900,000 calls gathered from a randomized controlled experiment in a live system, we showed that there is a strong bias in 
response rate due to the presentation order of questions. The most dramatic impact is experienced by the top-most token. In mobile environments, 
scrolling can lead to a reduction in response rate by as much as $49\%$. Motivated by these observations, we studied the problem of selecting 
a subset of tokens that maximize information while minimizing correlation. We achieved this by mapping it to the problem of submodularity maximization 
studied in the machine learning community. By doing so, we were able to retain $94\%$ of the information using just $30\%$ of the questions. Finally, 
we would like to emphasize that these methods and results can vastly benefit the media community as they significantly improve the quality of 
data gathered from any QoE survey.



\end{document}